\documentclass[aps,prd,twocolumn,superscriptaddress,longbibliography,nofootinbib]{revtex4-2}
\usepackage[utf8]{inputenc}
\usepackage{color}
\usepackage{graphicx}
\usepackage{subfigure}
\usepackage{multirow}

\usepackage{amsmath,amssymb,amsfonts}

\def\prl{Phys. Rev. Lett.}
\def\prd{Phys. Rev. D}

\def\pau_p{Prog. Theor. Phys.}

\begin{document}

\title{Bona-Masso slicing conditions and the lapse close to black-hole punctures}

\author{Thomas W.~Baumgarte}

\affiliation{Department of Physics and Astronomy, Bowdoin College, Brunswick, ME 04011, USA}

\author{Henrique P. de Oliveira}

\affiliation{Department of Physics and Astronomy, Bowdoin College, Brunswick, ME 04011, USA}
\affiliation{Departamento de F\'isica Te\'orica, Instituto de F\'isica A. D. Tavares, Universidade do Estato do Rio de Janeiro, R. S\~ao Francisco Xavier, 524, 20550-013, Rio de Janeiro, RJ, Brazil}

\begin{abstract}
We consider several families of functions $f(\alpha)$ that appear in the Bona-Masso slicing condition for the lapse function $\alpha$.  Focusing on spherically symmetric and time-independent slices we apply these conditions to the Schwarzschild spacetime in order to construct analytical expressions for the lapse $\alpha$ in terms of the areal radius $R$.  We then transform to isotropic coordinates and determine the dependence of $\alpha$ on the isotropic radius $r$ in the vicinity of the black-hole puncture.  We propose generalizations of previously considered functions $f(\alpha)$ for which, to leading order, the lapse is proportional to $r$ rather than a non-integer power of $r$.  We also perform dynamical simulations in spherical symmetry and demonstrate advantages of the above choices in numerical simulations employing spectral methods.  
\end{abstract}

\maketitle

\section{Introduction}
\label{sec:intro}

Among the most successful slicing conditions in numerical relativity is the Bona-Masso condition
\begin{equation} \label{slice_general}
    (\partial_t - \beta^i \partial_i) \, \alpha = - \alpha^2 f(\alpha) K,
\end{equation}
where $\alpha$ is the lapse function, $\beta^i$ the shift vector, and $K$ the trace of the extrinsic curvature (see \cite{BonMSS95}). Specific choices for the function $f(\alpha)$ single out specific slicing conditions; choosing $f(\alpha) = 1$, for example, results in harmonic slicing.  A very common choice for $f(\alpha)$ is the ``1+log" condition
\begin{equation} \label{1+log}
    f(\alpha) = \frac{2}{\alpha},
\end{equation}
which, together with a ``Gamma-driver" condition for the shift (e.g.~\cite{AlcB01,AlcBDKPST03}), forms the so-called ``moving-puncture" gauge conditions that have played a crucial role in simulations of black-hole spacetimes (see, e.g., \cite{CamLMZ06,BakCCKM06}).

A number of authors have suggested modifications to the Bona-Masso conditions and alternatives to the 1+log condition, for example to address the appearance of coordinate shocks \cite{Alc97}, to explore singularity avoidance \cite{Alc03}, or to improve the behavior of numerical simulations in the presence of adaptive mesh refinement interfaces \cite{EtiBPKS14}.  In this short paper we explore alternative choices for $f(\alpha)$ from a ``local" perspective, namely regarding the behavior of the lapse function $\alpha$ in the vicinity of a black-hole puncture.  For the 1+log slicing condition (\ref{1+log}), for example, spherically symmetric and time-independent solutions satisfy
\begin{equation} \label{alpha_of_r_1+log}
    \alpha \propto r^{1.091}~~~~~~~~~~(r \rightarrow 0),
\end{equation}
where $r$ is the isotropic radius (see \cite{Bru09} as well as Section \ref{sec:1+log} below).\footnote{In the absence of the shift term in (\ref{slice_general}), time-independent solutions are maximally sliced (see \cite{HanHBGSO07,BauN07}) and, adopting (\ref{1+log}), result in $\alpha \propto r^{\sqrt{2}}$ as $r \rightarrow 0$, see \cite{Bau11}.}  The appearance of the non-integer exponent in (\ref{alpha_of_r_1+log}) means that the second radial derivative of the lapse function diverges at the black-hole puncture, for $r \rightarrow 0$.  For numerical simulations employing spectral methods, powers with non-integer exponents are also difficult to express in terms of the most common basis functions, and therefore lead to slow convergence.  

Motivated by these considerations we generalize treatments by, e.g., Hannam {\it et.al.}~\cite{HanHPBM07,HanHOBO08} to construct spherically symmetric and time-independent Bona-Masso slices of the Schwarzschild spacetime for a number of different functions $f(\alpha)$.  We then follow Br\"uegmann \cite{Bru09} and transform to isotropic coordinates in order to determine the functional dependence of the lapse $\alpha$ on the isotropic radius $r$.   We identify special choices for $f(\alpha)$ for which the lapse $\alpha$ becomes (approximately) proportional to the isotropic radius $r$, rather than some non-integer power of $r$.  We also perform numerical simulations using a spectral code that implements the Baumgarte-Shapiro-Shibata-Nakamura (BSSN) equations \cite{NakOK87,ShiN95,BauS98} in spherical symmetry and demonstrate advantages of these special choices.

\section{Basic equations}
\label{sec:eqs}

\subsection{Transformation to Bona-Masso slices}
\label{sec:slices}

We start with the metric for a Schwarzschild black hole in Schwarzschild coordinates,
\begin{equation}
    ds^2 = - f_0 dt^2 + f_0^{-1} dR^2 + R^2 d\Omega^2.
\end{equation}
Here $R$ is the areal radius, we have defined $f_0 \equiv 1 - 2M/R$ where $M$ is the black hole mass, and we have adopted geometrized units with $c = G = 1$.  We transform to a new time coordinate $\bar t$ that is related to the old time coordinate $t$ by a ``height function" $h(R)$,
\begin{equation}
    \bar t = t + h(R)
\end{equation}
In terms of $\bar t$, the metric now takes the form
\begin{equation} \label{metric_t_bar}
    ds^2 = - f_0 d\bar t^2 + 2 f_0 h' d\bar t dR + (f_0^{-1} - f_0 h'^2) \, dR^2 + R^2 d \Omega^2
\end{equation}
where the prime denotes a derivative with respect to $R$, i.e.~$h' \equiv dh/dR$ (see, e.g., \cite{Rei73,BeiO98,MalO03}; see also Section 4.2 in \cite{BauS10} for a textbook treatment).  From (\ref{metric_t_bar}) we can identify the lapse function
\begin{equation}
    \alpha^2 = \frac{f_0}{1 - f_0 h'^2},
\end{equation}
the shift vector
\begin{equation} \label{shift}
    \beta^R = \frac{f_0^2 h'}{1 - f_0^2 h'^2} 
    = \alpha \, (\alpha^2 - f_0)^{1/2} ,
\end{equation}
and the $RR$-component of the spatial metric $\gamma_{ij}$
\begin{equation}
    \gamma_{RR} = \frac{1 - f_0^2h'^2}{f_0} = \alpha^{-2}.
\end{equation}
Finally, the trace of the extrinsic curvature can be written as 
\begin{align} \label{K1}
    K & = \frac{1}{R^2} \frac{d}{dR} \left(R^2 f_0 \alpha h' \right)
    = \frac{1}{R^2} \frac{d}{dR} \left( R^2 \frac{\beta^R}{\alpha} \right) \nonumber \\
    & = \frac{2}{R} \frac{\beta^R}{\alpha} + \frac{(\beta^{R})'}
    {\alpha}
    - \frac{\beta^R}{\alpha^2} \, \alpha'.
\end{align}

In spherical symmetry, and for time-independent slices, the slicing condition (\ref{slice_general}) results in
\begin{equation} \label{slice}
\beta^R \alpha' = \alpha^2 f(\alpha) K.
\end{equation}
We now follow \cite{HanHPBM07,HanHOBO08} and construct Bona-Masso slices by inserting (\ref{K1}) into (\ref{slice}) to obtain
\begin{equation}
    \frac{d\alpha}{\alpha f(\alpha)} + \frac{d \alpha}{\alpha} 
    = \frac{2 dR}{R} + \frac{d \beta^R}{\beta^R}.
\end{equation}
Integration then yields
\begin{equation} \label{first_integral}
    \alpha^2 = 1 - \frac{2M}{R} + \frac{C e^{2 I(\alpha)}}{R^4},
\end{equation}
where we have used (\ref{shift}), where $C$ is a constant of integration, and where $I(\alpha)$ is defined by the integral
\begin{equation} \label{I}
    I(\alpha) \equiv \int^\alpha_0 \frac{d \tilde \alpha}{\tilde \alpha f(\tilde \alpha)}.
\end{equation}

In order to determine the constant of integration in (\ref{first_integral}) we insert (\ref{shift}) into (\ref{K1}), and then use the result in (\ref{slice}) to obtain an equation for the derivative of the lapse $\alpha$ alone,
\begin{equation} \label{alpha_prime}
    \alpha' = - \frac{\alpha f(\alpha)}{R^2} \, \frac{3 M - 2 R + 2 R \alpha^2}{1 - 2 M/R + \alpha^2 f(\alpha) - \alpha^2}.
\end{equation}
We now observe that, for $\alpha'$ to remain regular across any point at which the denominator on the right-hand side of (\ref{alpha_prime}) vanishes, the numerator has to vanish simultaneously (see \cite{HanHPBM07}).  We refer to such a point as a ``critical point", and label the corresponding variables with a subscript $c$.  

For a given choice of the function $f(\alpha)$ we may therefore construct static and spherically symmetric black-hole slices as follows.  We first insert $f(\alpha)$ into (\ref{alpha_prime}) and search for simultaneous roots of the numerator and denominator, which, if they exists, determine the critical values $\alpha_c$ and $R_c$.  We then insert these values into (\ref{first_integral}), which determines the constant of integration $C$.  

We will be particularly interested in the behavior of the slices in the neighborhood of roots of the lapse.  Towards that end, we first determine the point $R_0$ at which $\alpha$ vanishes by evaluating (\ref{first_integral}),
\begin{equation}
    1 - \frac{2M}{R_0} + \frac{C e^{2 I(0)}}{R_0^4} = 0.
\end{equation}
While it may be possible to express solutions to this quartic equation in closed form, these expressions are often unwieldy and not particularly useful.  It is possible, however, to find numerical values for $R_0$ for some choices of the function $f(\alpha)$.

Finally, we take derivatives of the function (\ref{first_integral}) in order to evaluate
\begin{equation}
    a_1 \equiv \left( \frac{d\alpha}{dR} \right)_{R = R_0},
\end{equation}
in terms of which we may write the lapse in the neighborhood of its root as
\begin{equation}
    \alpha(R) = a_1 (R - R_0) + \mathcal{O}\left( (R - R_0)^2 \right)
\end{equation}
(see \cite{Bru09}).

\subsection{Transformation to isotropic coordinates}
\label{sec:isotropic}

We now transform the spatial metric on slices of constant time $\bar t$ to isotropic coordinates with a radial coordinate $r$.  In particular, we identify the spatial line element corresponding to the spacetime line element (\ref{metric_t_bar}),
\begin{equation}
    dl^2 = \alpha^{-2} dR^2 + R^2 d\Omega^2,
\end{equation}
with that of a spatial metric in isotropic coordinates,
\begin{equation}
    dl^2 = \psi^4 ( dr^2 + r^2 d\Omega^2),
\end{equation}
where $\psi$ is a conformal factor.  This identification results in the two conditions
\begin{equation}
\alpha^{-1} dR = \psi^2 dr~~~~~\mbox{and}~~~~~R = \psi^2 r, 
\end{equation}
which we may combine to obtain
\begin{equation}
\frac{dr}{r} = \frac{dR}{R \alpha} = \frac{dR/d\alpha}{R} \, \frac{d \alpha}{\alpha}
\end{equation}
and hence
\begin{equation} \label{r1}
    r = \exp \int \frac{dR / d\alpha}{R} \, \frac{d \alpha}{\alpha}. 
\end{equation}
To leading order in $r$ we may now approximate $dR / d\alpha \simeq 1 / a_1$ and $R \simeq R_0$ and carry out the integration to obtain
\begin{equation} \label{r2}
    r \propto \alpha^\gamma~~~~~~~~~~(r \rightarrow 0),
\end{equation}
where, following \cite{Bru09}, we have defined
\begin{equation} \label{gamma}
    \gamma \equiv \frac{1}{a_1 R_0}.
\end{equation}
Inverting (\ref{r2}) we find that, in the vicinity of the black-hole puncture, the lapse behaves according to
\begin{equation} \label{alpha_power}
    \alpha \propto r^{1/\gamma}~~~~~~~~~~(r \rightarrow 0)
\end{equation}
({\it cf.}~Eq.~(56) in \cite{Bru09}).

\section{Examples and generalizations}
\label{sec:examples}

In this Section we consider several examples for the function $f(\alpha)$, and suggest generalizations that result in exponents $1/\gamma \simeq 1$.  We summarize our findings in Table~\ref{tab:summary}.

\begin{table*}[t]
    \centering
    \begin{tabular}{c|c|c|c|c|c|c|c|c}
         $f(\alpha)$ & Ref. & $I(\alpha)$ & parameter & $\alpha_c$ & $R_c / M$ & $C / M^4$ & $R_0 / M$ & $1 / \gamma$  \\
         \hline \hline
        \multirow{2}{*}{$\displaystyle k / \alpha $} & 
        \multirow{2}{*}{\cite{BonMSS95}} & \multirow{2}{*}{$\alpha / k$} & $k = 2$ & 0.162 & 1.541 & 1.554 & 1.312 & 1.091 \\
        & &  & $k = 1.46263$ & 0.217 & 1.574 & 1.450 & 1.240 & 1.000 \\
        \hline
        $(1-\alpha)/\alpha$ & \cite{DenB14} &  $- \ln(1 - \alpha)$ & -- & 1/2 & 2 & 1 & 1 & 1 \\
        \hline
                $\displaystyle 1 + \frac{\kappa}{\alpha^2}$ &
        \cite{Alc97} &
        $\displaystyle \frac{1}{2}\ln \left( \frac{\alpha^2 + \kappa}{\kappa} \right)$ & $\kappa > 1/3$ & 0 & $3/2$ & $3^3/2^4$ & 3/2 &
        $\displaystyle \left( \frac{6 \kappa}{3\kappa - 1} \right)^{1/2}$ \\
        \hline
        \multirow{2}{*}{$\displaystyle \frac{a_0^2}{2 \alpha + (a_0 - 2) \alpha^2}$} & 
        \multirow{2}{*}{\cite{Alc03}} &
        \multirow{2}{*}{$\displaystyle \frac{\alpha}{2 a_0^2} \Big( 4 + ( a_0 - 2 ) \alpha \Big)$} &
        $a_0 = 4/3$ & 0.305 & 1.654 & 1.179 & 1.090 & 0.801 \\
        & & &  $a_0 = 1.7365$ & 0.206 & 1.567 & 1.468 & 1.252 & 1.000
    \end{tabular}
    \caption{A summary of our results for different families of the Bona-Masso functions $f(\alpha)$.  For each family we list the integral (\ref{I}), and, for selected parameter choices, the critical values of the lapse $\alpha_c$ and areal radius $R_c$, the integration constant $C$ in (\ref{first_integral}), the areal radius $R_0$ at which the lapse vanishes, and the exponent $1 / \gamma$ in (\ref{alpha_power}) that determines the power-law behavior in the vicinity of the black-hole puncture, $\alpha \propto r^{1/\gamma}$. }
    \label{tab:summary}
\end{table*}

\subsection{1+log slicing}
\label{sec:1+log}

The 1+log slicing condition \cite{BonMSS95} is obtained for the choice (\ref{1+log}).  Integration of (\ref{I}) then yields
\begin{equation}
    I(\alpha) = \frac{\alpha}{2},
\end{equation}
so that (\ref{first_integral}) becomes
\begin{equation} \label{first_integral_1+log}
\alpha^2 = 1 - \frac{2M}{R} + \frac{C e^{\alpha}}{R^4}
\end{equation}
(see \cite{HanHPBM07}).  From the simultaneous roots of the numerator and denominator of (\ref{alpha_prime}) we then find
\begin{subequations}
\begin{align}
    \alpha_c & = \sqrt{10} - 3 \simeq 0.162 \\
    R_c & = \frac{3 + \sqrt{10}}{4} \, M \simeq 1.541 \, M,
\end{align}
\end{subequations}
which, when inserted into (\ref{first_integral_1+log}), yields
\begin{equation}
    C = \frac{1}{128}\left(3 + \sqrt{10} \right)^3 e^{3 - \sqrt{10}} \simeq 1.554 \, M^4
\end{equation}
(see also \cite{HanHOBO08}).  We next evaluate (\ref{first_integral_1+log}) at $\alpha = 0$ to find
\begin{equation}
    R_0 \simeq 1.312 \, M
\end{equation}
(another real root of $R_0 \simeq 1.66$ leads to negative values of $\alpha$ for $R > R_0$), as well as
\begin{equation}
    a_1 \simeq 0.832 \, M^{-1}.
\end{equation}
Finally, we compute $\gamma \simeq 0.916$ from (\ref{gamma}), so that 
\begin{equation}
    \alpha \propto r^{1.091}.
\end{equation}
All of the above is in complete agreement with the results of \cite{Bru09}.

As a generalization of (\ref{1+log}) we may consider
\begin{equation} \label{1+log_general}
    f(\alpha) = \frac{k}{\alpha}
\end{equation}
(see, e.g., \cite{HanHOBO08}).  Carrying out the same calculations as above we find that, for 
\begin{equation}
    k \simeq 1.46263,
\end{equation}
we obtain $\gamma = 1.0$ to high accuracy, so that the lapse $\alpha$ is approximately proportional to the isotropic radius $r$ close to the black-hole puncture (see Table \ref{tab:summary} for details).

\subsection{Analytical trumpet slices}
\label{sec:analytical}

As an alternative we consider
\begin{equation} 
    f(\alpha) = \frac{1 - \alpha}{\alpha},
\end{equation}
which results from the analytical trumpet slices constructed in \cite{DenB14}.\footnote{This choice is the special case $R_0 = M$ of a larger family satisfying $f(\alpha) = (1 - \alpha) \alpha^{-1} (2 M - R_0 (1 + \alpha)) / (3 M - R_0 (2 + \alpha))$; see \cite{DenB14} for details.}  We can again integrate (\ref{I}) analytically,
\begin{equation}
    I(\alpha) = - \ln(1 - \alpha),
\end{equation}
so that (\ref{first_integral}) becomes
\begin{equation} \label{first_integral_analytical}
    \alpha^2 = 1 - \frac{2M}{R} + \frac{C^2}{R^4}\, \frac{1}{(\alpha - 1)^2}.
\end{equation}
The numerator and denominator of (\ref{alpha_prime}) now have simultaneous roots for
\begin{equation}
    \alpha_c = \frac{1}{2},~~~~~~~R_c = 2 \, M,
\end{equation}
which, when inserted into (\ref{first_integral_analytical}), yields $C = M^4$.  The desirable root of (\ref{first_integral_analytical}) for $\alpha = 0$ is $R_0 = M.$ We then have $a_1 = M^{-1}$ and therefore $\gamma = 1$ exactly, indicating that the lapse now satisfies
\begin{equation}
    \alpha \propto r
\end{equation}
close to the origin, in complete agreement with the analytical solution
\begin{equation}
    \alpha = \frac{r}{r + M}
\end{equation}
provided in \cite{DenB14}.  We note, however, that $f(\alpha) \rightarrow 0$ as $\alpha \rightarrow 1$, making this choice undesirable in general (in spherical symmetry,  however, it provides a powerful numerical test with a simple analytical solution).  


\subsection{Gauge-shock avoiding slices}
\label{sec:gauge_shock}

As a means to avoid gauge shocks, Alcubierre \cite{Alc97} suggested
\begin{equation} \label{gauge_shock_f}
    f(\alpha) = 1 + \frac{\kappa}{\alpha^2}, 
\end{equation}
as yet another alternative choice for the function $f(\alpha)$ (see also \cite{Alc03} as well as \cite{JimVA21} for numerical simulations with $\kappa = 1$).  The integral (\ref{I}) can again be carried out analytically,
\begin{equation} \label{gauge_shock_I}
I(\alpha) = \frac{1}{2} \ln\left( \frac{\alpha^2 + \kappa}{\kappa} \right),
\end{equation}
where we have assumed $\kappa > 0$.  

This case differs from the previous cases, however, in that $\alpha f(\alpha)$ on the right-hand side of (\ref{alpha_prime}) diverges as $\kappa / \alpha$ as $\alpha \rightarrow 0$.  Therefore, a root of the denominator of (\ref{alpha_prime}) may result from a vanishing of $\alpha$ rather than a root of the denominator of the second fraction on the right-hand side of (\ref{alpha_prime}).  In fact, for $\kappa > 1/3$, the outermost root of the denominator (i.e.~the one for the largest radius) occurs for $\alpha = 0$ so that $\alpha_c = 0$.  The critical radius $R_c$ is hence equal to $R_0$ and takes the value $R_c = 3 M /2$.  Inserting (\ref{gauge_shock_I}) into (\ref{first_integral}) we find $C = 3^3 M^4 / 2^4$ as well as $a_1$, and finally
\begin{equation}
    \frac{1}{\gamma} = \left( \frac{6 \kappa}{3 \kappa - 1} \right)^{1/2}~~~~~~~~~~~~~(\kappa > 1/3).
\end{equation}
For $\kappa = 1$, for example, we have $1/\gamma = \sqrt{3}$.  However, we can also make $1/\gamma$ take an integer value $n$ by choosing $\kappa = n^2 /(3n^3 - 6)$.  An attractive choice from our perspective here, while satisfying our assumption $\kappa > 1/3$, is $\kappa = 2/3$, which results in $1/\gamma = 2$. 

In numerical experiments with (\ref{gauge_shock_f}), however, we found that the lapse can take negative values during the evolution, as anticipated in \cite{Alc03}, and dynamical evolutions also appear to take significantly longer to settle down to equilibrium than for the other choices of $f(\alpha)$ discussed here.  We therefore follow \cite{Alc03} in considering functions $f(\alpha)$ that are gauge-shock avoiding to leading order only.

Specifically, we adopt the ansatz (82) of \cite{Alc03}, which is defined in terms of coefficients $p_0$, $q_1$, and $q_2$.  Rather than imposing first-order shock avoidance, which results in the conditions (83)--(85) of \cite{Alc03} for these coefficients, we require that $f(\alpha) \propto \alpha^{-1}$ as $\alpha \rightarrow 0$, which instead results in the condition $q_2 = q_1 - 1$.  Further imposing zeroth-order shock avoidance then leads to the family
\begin{equation} \label{zero_order_f}
    f(\alpha) = \frac{a_0^2}{2 \alpha + (a_0 - 2) \alpha^2},
\end{equation}
where the parameter $a_0 = q_0$ yields the value of $f(\alpha)$ for $\alpha = 1$.  Note that the 1+log slicing condition (\ref{1+log}) is a member of this family with $a_0 = 2$, while first-order shock avoidance is achieved for $a_0 = 4/3$ (see \cite{Alc03} as well as \cite{RucHLZ17} for numerical experiments).  

For (\ref{zero_order_f}), the integral (\ref{I}) can be evaluated to yield
\begin{equation}
    I(\alpha) = \frac{\alpha}{2 a_0^2} \left( 4 + (a_0 - 2) \alpha \right),
\end{equation}
so that (\ref{first_integral}) becomes 
\begin{equation} \label{first_integral_gauge_shock}
\alpha^2 = 1 - \frac{2M}{R} + \frac{C^2 \exp\left\{\alpha (4 + (a_0 - 2) \alpha)/a_0^2 \right\}}{R^4}.
\end{equation}
Adopting $a_0 = 4/3$ we find $\alpha_c \simeq 0.305$ and $R_c \simeq 1.654 \, M$ from simultaneous roots of the numerator and denominator of (\ref{alpha_prime}), which, when inserted into (\ref{first_integral_gauge_shock}), yields $C \simeq 1.179 \, M^4$.  A root of (\ref{first_integral_gauge_shock}) is then given by $R_0 \simeq 1.090 \, M$, from which we compute $\gamma = 1.249$ and hence
\begin{equation}
    1 / \gamma \simeq 0.801 ~~~~~~~~~~~~(a_0 = 4/3).
\end{equation}
Repeating the analysis for 
\begin{equation}
    a_0 = 1.7365, 
\end{equation}
however, we find $1/\gamma$ very close to unity, so that the lapse is again approximately proportional to the isotropic radius $r$ in the vicinity of the black-hole puncture.

\section{Numerical examples}
\label{sec:numerics}

We next present numerical examples in order to illustrate some of the results of Section \ref{sec:examples}, and to demonstrate the respective advantages and disadvantages of some of the choices.  We will focus on single black holes in spherical symmetry, evolving the BSSN equations with spectral methods in the context of the moving-puncture method without excision.


\subsection{Spectral code}
\label{sec:spectral}

Our code solves the BSSN equations in spherical symmetry (see, e.g., \cite{MonC12}) using a multi-domain Galerkin-collocation spectral method (\cite{AlcBO21,AlcABO21}, see also \cite{Boy01} for a textbook treatment).  Details of this code will be presented elsewhere (see \cite{Oli22}), so that we will discuss only some of its main features here.  

Rather than using the radius $r$, our code uses a coordinate $x = L(r) = (r-L_0)/(r+L_0)$, where $L(r)$ maps the infinite domain $(0,\infty)$ into the finite domain $(-1,1)$, and where $L_0$ is a parameter with dimension of length.  Our code also allows this ``global" domain to be split into multiple sub-domains; for the examples presented below we will use two such sub-domains.  In each sub-domain we apply a second, linear map so that the local coordinates again cover the interval $(-1,1)$, and then expand all functions into basis functions.  For the lapse function $\alpha(t,r)$, for example, we write
\begin{equation} \label{expansion}
\alpha(t,x) = 1 + \sum_{k=0}^N\,\hat{\alpha}_k(t) \psi_k(x)
\end{equation}
in each sub-domain, where the $\hat \alpha_k(t)$ are mode coefficients, the $\psi_k(x)$ form a complete set of basis functions, and where $N$ is the truncation order.  In the inner sub-domains we adopt the rational Chebyshev functions $\psi_k(x) = T_k(x)$ as basis functions (see, e.g., \cite{Boy01}), while, in the outermost sub-domain, we use the combinations $\psi_k(x) = T_{k+1}(x) - T_k(x)$, since the latter automatically satisfy the boundary conditions at spatial infinity.

We next insert the expansion for all dynamical fields into the BSSN equations, the slicing condition (\ref{slice_general}) for the lapse, and the ``Gamma-driver" gauge condition for the shift (see \cite{AlcB01,AlcBDKPST03}).  Evaluating these equations at $N+1$ collocation points then casts the set of coupled partial differential equations as a set of coupled ordinary differential equations for the mode coefficients.  We integrate this set of equations using a standard Runge-Kutta method, and thus obtain all mode coefficients, e.g.~the $\hat \alpha_k(t)$, as functions of time.  Finally we can reconstruct the physical fields by inserting these mode coefficients into the respective expansions, e.g.~(\ref{expansion}).  We again refer to \cite{Oli22} for a more detailed description and discussion.

\subsection{Numerical results}
\label{sec:results}

We adopt ``wormhole" data as initial data, i.e.~the Schwarzschild solution in isotropic coordinates on a slice of constant Schwarzschild time.  At the initial time we also choose a ``pre-collapsed" lapse, $\alpha = (1 + M/(2r))^{-2}$, as well as zero shift.  We then evolve these data with the Bona-Masso slicing condition (\ref{slice_general}) for two families of functions $f(\alpha)$, namely the ``generalized 1+log" slices of Section \ref{sec:1+log} (see Eq.~\ref{1+log_general}) and the ``zeroth-order gauge-shock avoiding" slices of Section \ref{sec:gauge_shock} (see Eq.~\ref{zero_order_f}), both for different choices of the respective parameters $k$ and $a_0$. For all results shown here we use two sub-domains, and 90 collocation points in each sub-domain.


\begin{figure}[t]
\centering
\includegraphics[width = 0.45 \textwidth]{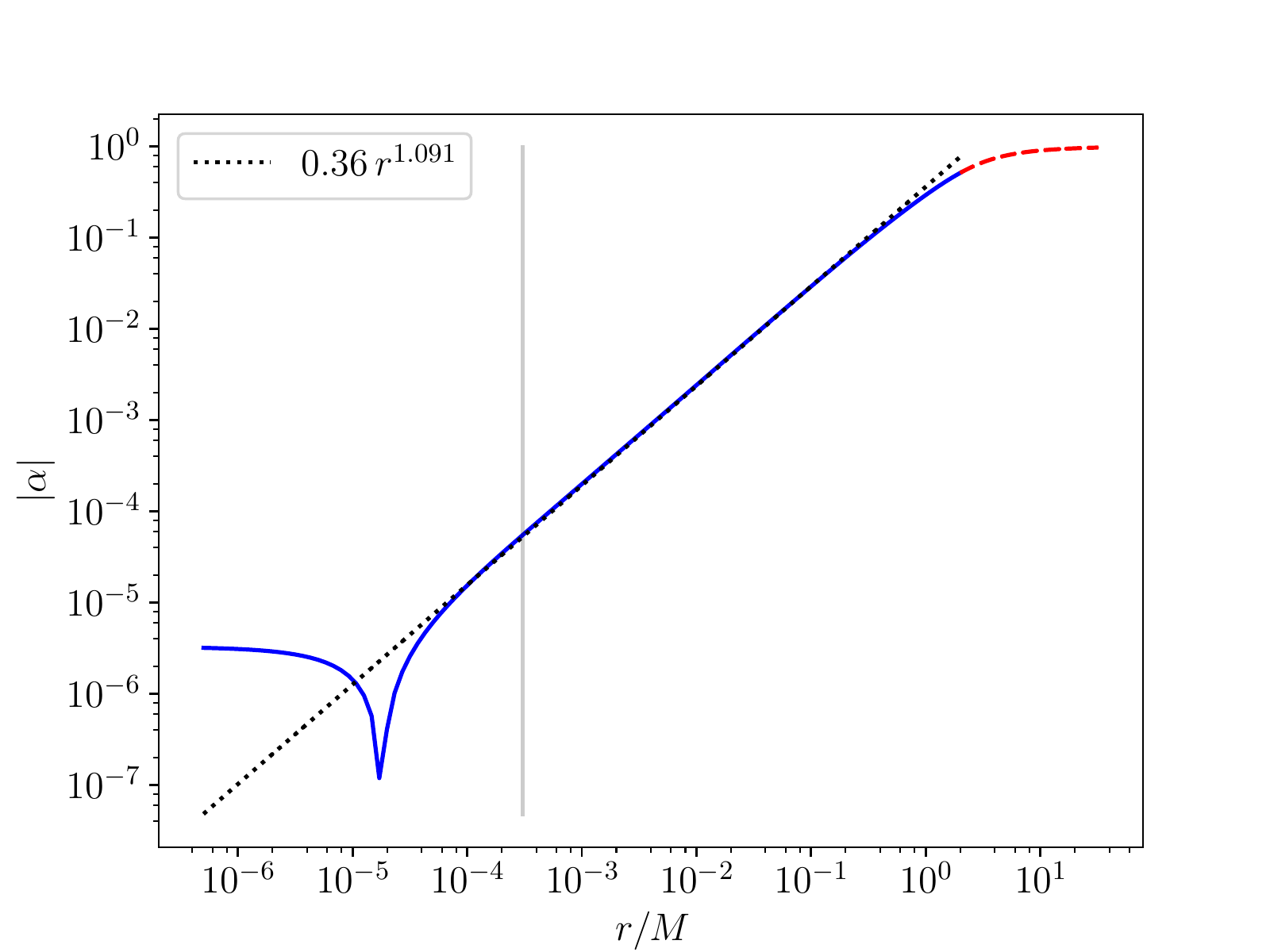}\\
\includegraphics[width = 0.45 \textwidth]{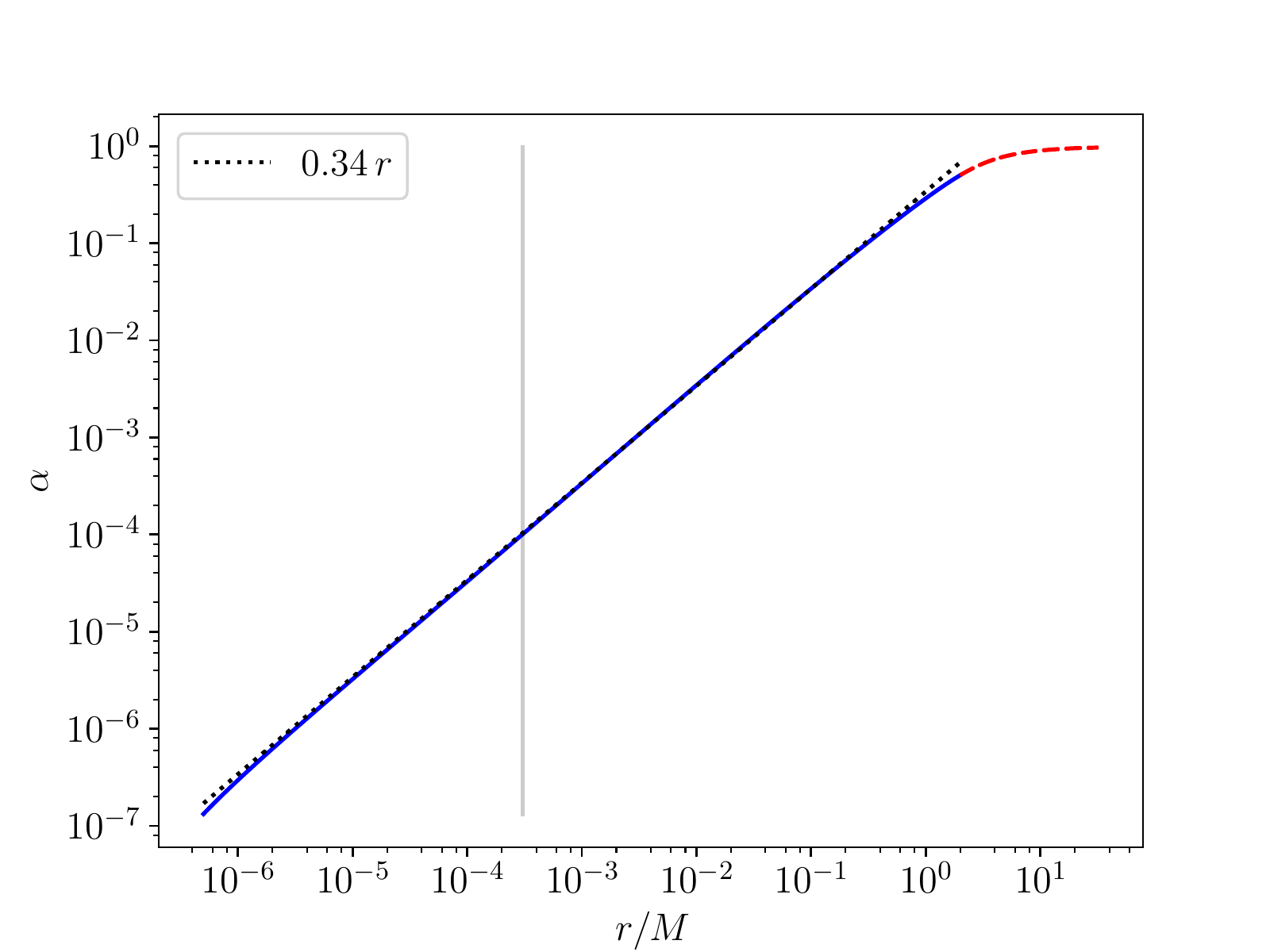}\\
	\caption{Profiles of the lapse function $\alpha$ as a function of radius $r$ at a time $t = 50 M$ for $f(\alpha) = k / \alpha$ (see Eq.~\ref{1+log_general}).  The top panel shows results for $k = 2$, and the bottom panel for $k=1.46263$.  The solid (blue) and dashed (red) lines show numerical results in the inner and outer sub-domain, while the dotted line shows the expected power-law scaling $\alpha \propto r^{1/\gamma}$, with $1/\gamma \simeq 1.091$ for $k = 2$ (top panel) and $1/\gamma \simeq 1$ for $k=1.46263$.  The vertical (grey) line marks the location of the innermost collocation point at $r_{\rm inner} \simeq 3.05 \times 10^{-4} M$.  Note that, for $k=1.46263$, the numerical solution follows the expected power-law to much smaller values of $r$ than for $k = 2$.
 }
\label{fig:lapse_1+log}
\end{figure}

During the first phase of the numerical evolution the fields change with time as the data undergo a coordinate transition from the initial wormhole geometry to a trumpet geometry (see, e.g., \cite{HanHOBO08}).  This transition takes a time of approximately $30 M$ or so, after which the evolution settles down into a new equilibrium and the data become approximately time-independent (at least in the vicinity of the black hole).  In the following we show results from our dynamical evolutions at a time $t = 50M$, when the data should be well approximated by the equilibrium solutions constructed in Section \ref{sec:examples}.

We first consider the ``generalized 1+log" choice $f(\alpha) = k/\alpha$ of Section \ref{sec:1+log}.  In Fig.~\ref{fig:lapse_1+log} we show profiles of the lapse $\alpha$ as a function of radius $r$ at $t = 50M$, both for the ``canonical" choice $k = 2$ (top panel) and for $k = 1.46263$ (bottom panel).  In both panels we include the numerical results as solid (blue) and dashed (red) lines (in the inner and outer sub-domain), as well as the expected power-law scaling (\ref{alpha_power}) as dotted lines.  We also mark the location $r_{\rm inner}$ of the innermost collocation point by the vertical lines.  Evidently, the expected power-law scaling extends to much smaller radii for $k = 1.46263$, when the power-law exponent is approximately unity, than for $k = 2$, when the exponent takes a non-integer value.  In particular, we see that the numerical results reproduce the expected power-law behavior to $r \ll r_{\rm inner}$ in the former case, but only to about $r \simeq r_{\rm inner}$ in the latter case.

\begin{figure}[t]
\centering
\includegraphics[width = 0.45 \textwidth]{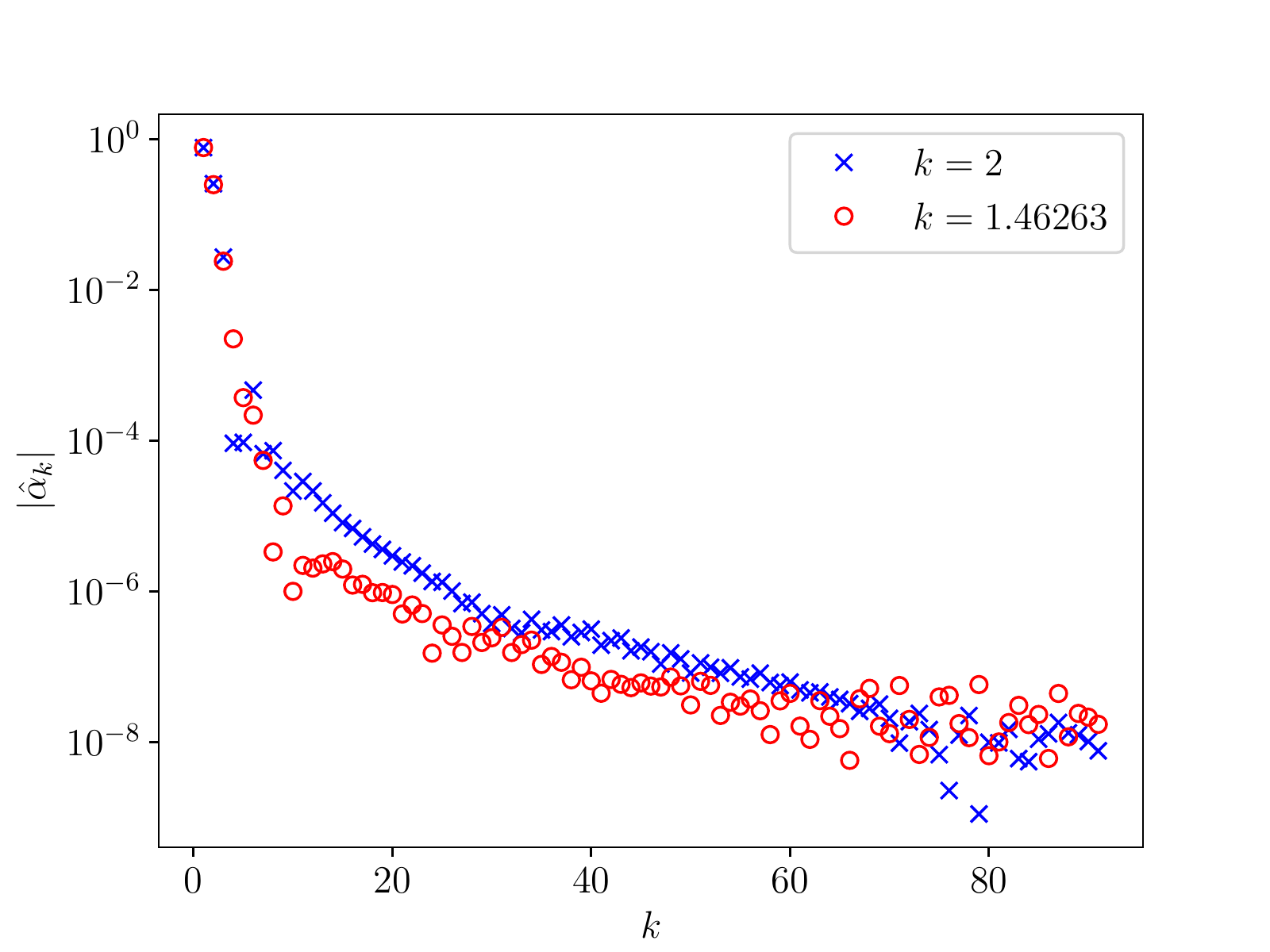}
	\caption{The mode coefficients $\hat \alpha_k$ at time $t = 50 M$ for $f(\alpha) = k /\alpha$ with $k=2$ (blue crosses) and $k=1.46263$ (red circles) in the inner sub-domain.} 
\label{fig:coefficients_1+log}
\end{figure}

The improved numerical behavior can also be seen in Fig.~\ref{fig:coefficients_1+log}, where we show the coefficients $\hat \alpha_k$ corresponding to the solutions shown in Fig.~\ref{fig:lapse_1+log} in the inner sub-domain.  We see that, for $k = 1.46263$, the coefficients (marked by red circles) drop off somewhat faster than for $k = 2$ (marked by blue crosses), and reach the noise level of approximately $10^{-7}$ (which is caused at least in part by deviations from true equilibrium) for smaller values of $k$.  While the difference is clearly noticeable, it is not very large, presumably because the expected power-law exponent of $1/\gamma \simeq 1.091$ for $k = 2$ is not very different from unity, the expected exponent for $k = 1.46263$.

\begin{figure}[t]
\includegraphics[width = 0.45 \textwidth]{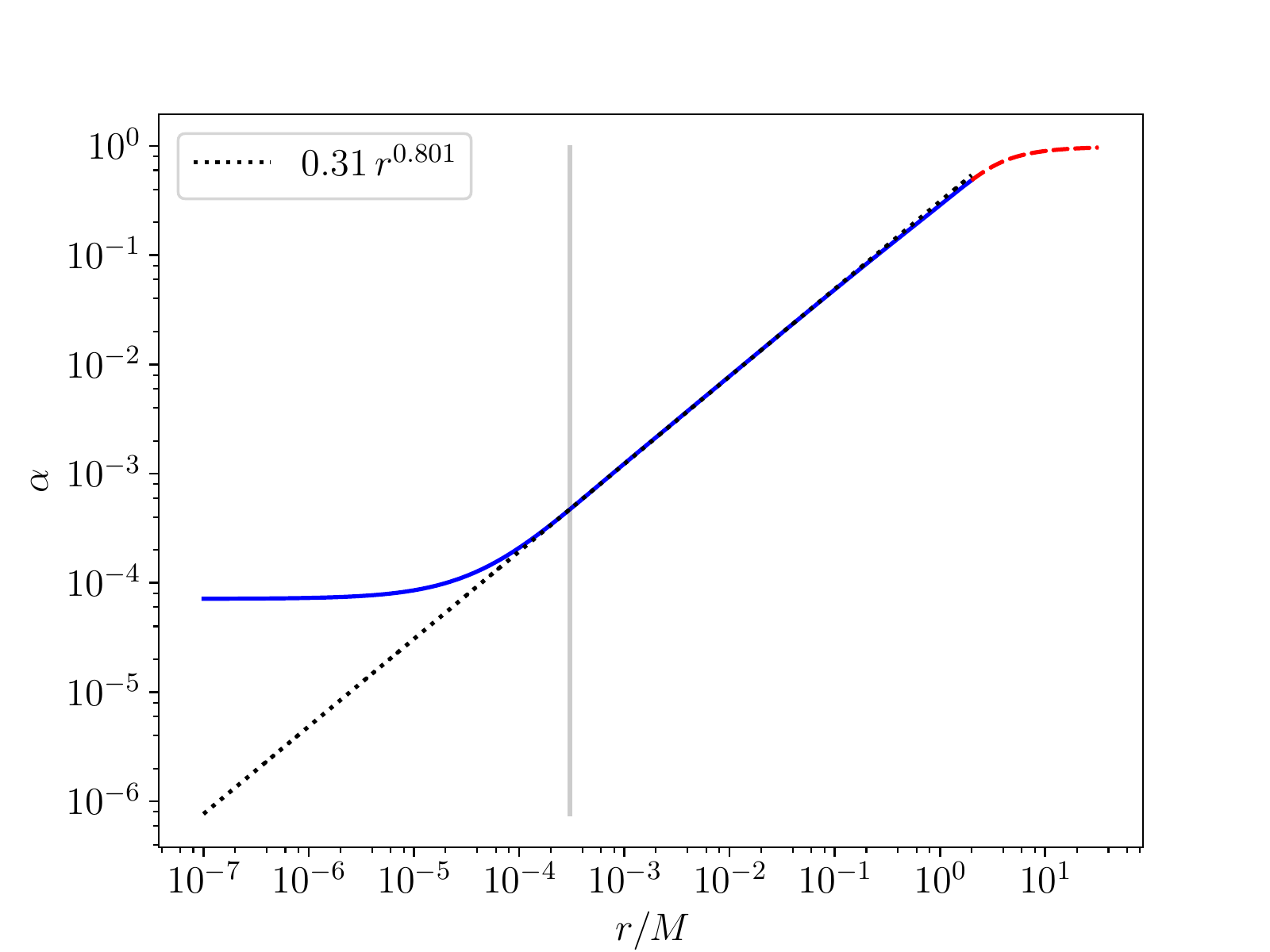}
\includegraphics[width = 0.45 \textwidth]{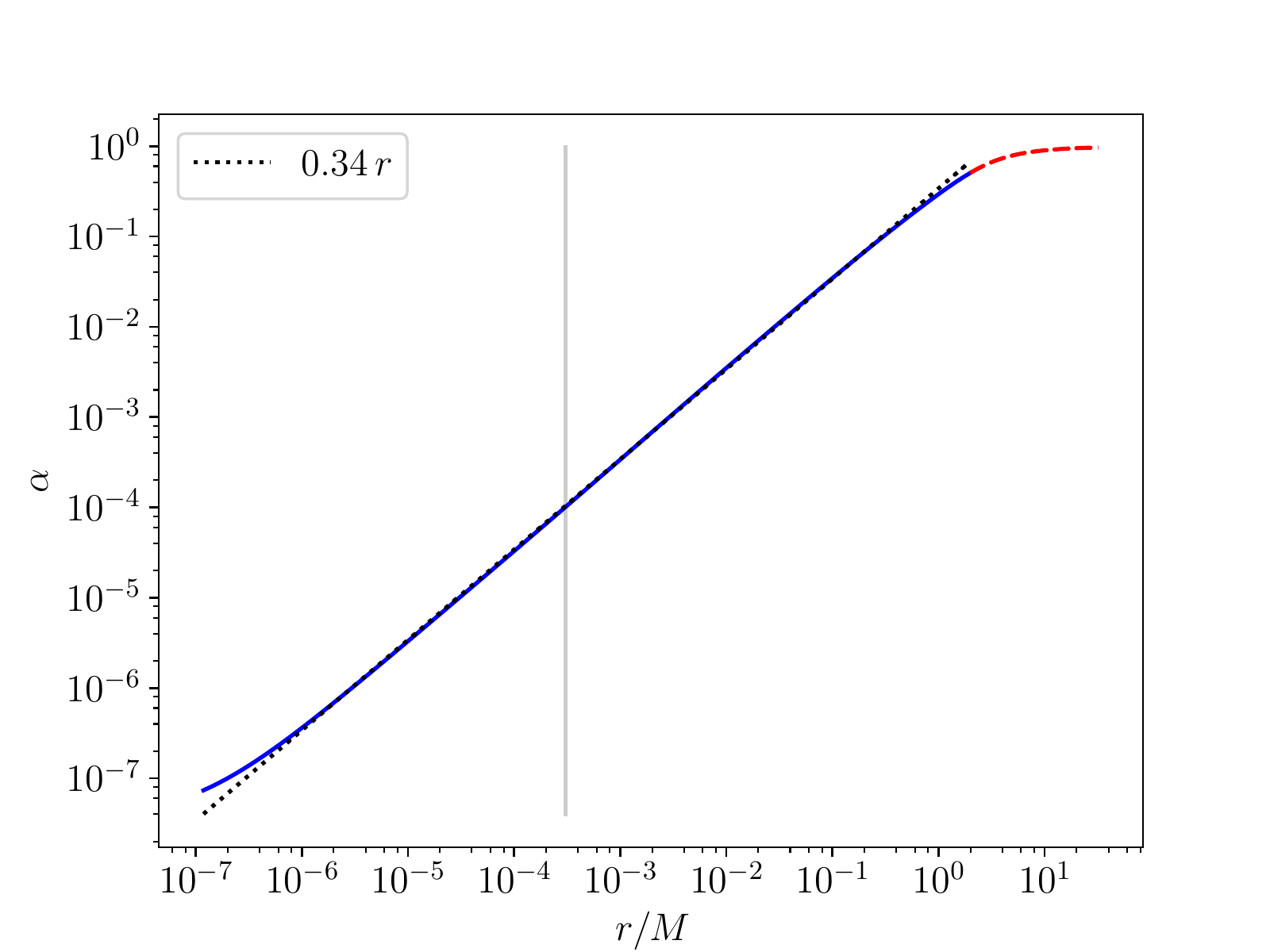}
	\caption{Same as Fig.~\ref{fig:lapse_1+log}, but for the function $f(\alpha)$ given by (\ref{zero_order_f}) with $a_0 = 4/3$ in the top panel and $a_0 = 1.7365$ in the bottom panel.  The dotted lines represent the expected power-law scalings $\alpha \propto r^{1/\gamma}$ with $1/\gamma \simeq 0.801$ in the top panel and $1/\gamma \simeq 1$ in the bottom panel. 
 } 
\label{fig:lapse_shock}
\end{figure}

\begin{figure}[t]
\centering
\includegraphics[width= 0.45 \textwidth]{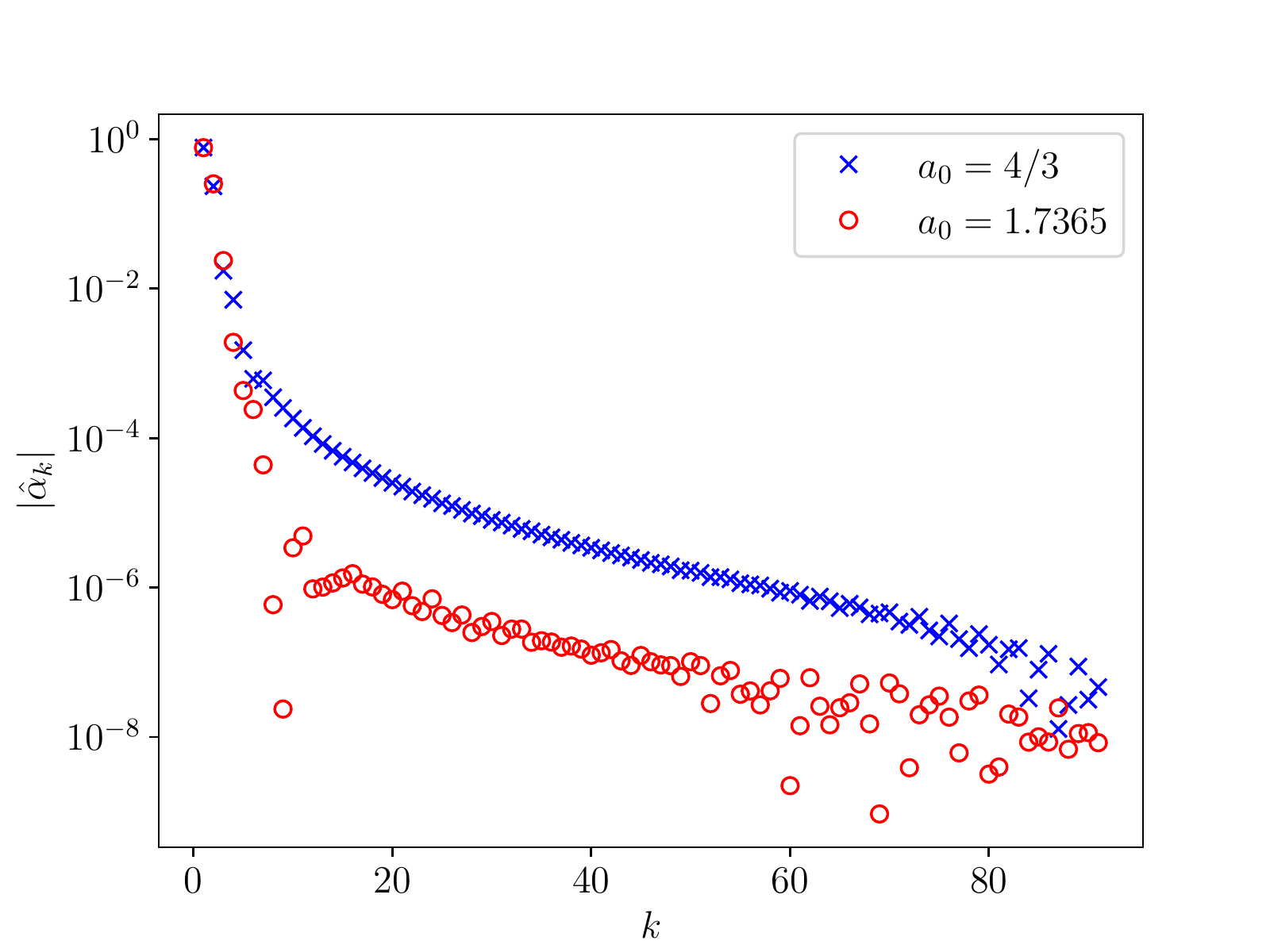}
	\caption{Same as Fig.~\ref{fig:coefficients_1+log} but for the family (\ref{zero_order_f}) with $a_0=4/3$ (blue crosses) and $a_0 = 1.7365$ (red circles).} 
\label{fig:coefficients_shock}
\end{figure}

In order to explore this behavior for a different example we next consider the family (\ref{zero_order_f}) with $a_0 = 4/3$ and $a_0 = 1.7365$.  In Fig.~\ref{fig:lapse_shock} we again show profiles of the lapse $\alpha$ as a function of radius $r$ at time $t = 50 M$.  As before, we observe much better agreement between the numerical solution and the expected power-law scaling at radii well inside the innermost collocation point when the exponent is (approximately) unity (for $a_0 = 1.7365$) than for a non-integer exponent (for $a_0 = 4/3$).  

This improvement is also evident in Fig.~\ref{fig:coefficients_shock}, where we show the corresponding mode coefficients $\hat \alpha_k$.  Clearly, these coefficients drop off much faster for $a_0 = 1.7365$ (red circles) than for $a_0 = 4/3$ (blue crosses), indicating that constructing the latter with an expansion (\ref{expansion}) requires many more basis functions $\psi_k$ than the former.  This is not unexpected, of course, since the latter features a non-integer power-law for small radii $r$, while, for the former, the behavior $\alpha \propto r$ can be reproduced quite easily in a spectral representation.  In our example we find that the coefficients drop off to a level of about $10^{-6}$ for $k \simeq 10$ for $a_0 = 1.7365$, but only for $k \simeq 60$ for $a_0 = 4/3$, indicating a much improved convergence for the former.  We believe that the improvement seen in this example is larger than that shown in Fig.~\ref{fig:coefficients_1+log} because here the difference in the power-law exponents is also larger: for $a_0 = 4/3$ we expect $1/\gamma \simeq 0.801$, which differs from unity more than the exponent $1/\gamma \simeq 1.091$ found for $k = 2$ in Fig.~\ref{fig:coefficients_1+log}.

\section{Summary}
\label{sec:summary}

We generalize the treatments of Hannam {\it et.al.}~\cite{HanHPBM07,HanHOBO08} and Br\"ugmann \cite{Bru09} to construct spherically symmetric and time-independent slices of the Schwarzschild spacetime satisfying the Bona-Masso slicing condition (\ref{slice_general}) for a number of different functions $f(\alpha)$.  Specifically, we derive analytical expressions for the lapse function $\alpha$ in terms of the areal radius $R$, and then transform these expressions to isotropic coordinates in order to obtain the leading-order dependence of $\alpha$ on the isotropic radius $r$ in the vicinity of the black-hole puncture, $\alpha \propto r^{1/\gamma}$ (see Eq.~\ref{alpha_power}).  

For many common choices of $f(\alpha)$, the exponent $1/\gamma$ takes non-integer values (see Table \ref{tab:summary}), which may have undesirable consequences for numerical simulations, in particular in the context of spectral methods.  We suggest generalizations of these functions $f(\alpha)$ for which $1/\gamma$ takes either exact or approximate integer values.

Finally, we perform numerical simulations using a spectral implementation of the BSSN equations in spherical symmetry.  Adopting ``wormhole" initial data we compare evolutions for different choices of $f(\alpha)$, and demonstrate the improved convergence for those functions $f(\alpha)$ that feature integer exponents $1/\gamma$.  

\acknowledgments

HPO acknowledges the financial support of the Brazilian Agency CNPq as well as hospitality at Bowdoin College and its Department of Physics and Astronomy. This work was supported in part by National Science Foundation (NSF) grant PHY-2010394 to Bowdoin College and the Coordena\c c\~ao de Aperfei\c coamento de Pessoal de N\'ivel Superior - Brasil (CAPES) - Finance Code 001.


%

\end{document}